\documentclass[a4paper,fleqn,usenatbib]{mnras}

\usepackage{newtxtext,newtxmath}

\usepackage[T1]{fontenc}
\usepackage{ae,aecompl}

\usepackage{graphicx}	
\usepackage{amsmath}	
\usepackage{amssymb,url}	

\def\simlt{\ \raise -2.truept\hbox{\rlap{\hbox{$\sim$}}\raise5.truept   %
		\hbox{$<$}\ }}
\def\simgt{\ \raise -2.truept\hbox{\rlap{\hbox{$\sim$}}\raise5.truept   %
		\hbox{$>$}\ }}                                                          %

\newcommand{\pd}[3]{\frac{\partial^{#3} #1}{\partial {#2}^{#3}}} 
\newcommand{\td}[3]{\frac{d^{#3} #1}{d {#2}^{#3}}} 
\renewcommand{\v}[1]{\ensuremath{\mathbf{#1}}} 
\newcommand{\gv}[1]{\ensuremath{\mbox{\boldmath$ #1 $}}}
\renewcommand{\bar}[1]{\ensuremath{\overline{#1}}}

\title[]{Hunting dark matter in galaxy clusters with non-thermal electrons}

\author[G. Beck]{
	Geoff Beck\thanks{E-mail: geoffrey.beck@wits.ac.za}
	\\
	School of Physics, University of the Witwatersrand, Johannesburg, WITS-2050, South Africa\\
}

\date{Accepted XXX. Received YYY; in original form ZZZ}

\pubyear{2019}

\begin{document}
\label{firstpage}
\pagerange{\pageref{firstpage}--\pageref{lastpage}}
\maketitle

\begin{abstract}
The electron population inferred to be responsible for the mini-halo within the Ophiuchus galaxy cluster is a steep power-law in energy with a slope of $3.8$. This is substantially different to that predicted by dark matter annihilation models. In this work we present a method of indirect comparison between the observed electron spectrum and that predicted for indirect dark matter emissions. This method utilises differences in the consequences of a given electron distribution on the subsequent spectral features of synchrotron emissions. To fully exploit this difference, by leveraging the fact that the peak and cut-off synchrotron frequencies are substantially different to hard power-law cases for WIMP masses above $\sim 50$ GeV, we find that we need $\mu$Jy sensitivities at frequencies above 10 GHz while being sensitive to arcminute scales. We explore the extent to which this electron spectrum comparison can be validated with the up-coming ngVLA instrument. We show that, with the ngVLA, this method allows us to produce far stronger constraints than existing VLA data, indeed these exceed the Fermi-LAT dwarf searches in a wide variety of annihilation channels and for all studied magnetic field scenarios.
\end{abstract}




\section{Introduction}
As highly Dark Matter (DM) dominated environments galaxy clusters might be expected to yield strong indirect DM signatures. However, gamma-ray searches have, so far, been fruitless~\citep{S.ZimmerfortheFermi-LAT:2015oka}. This has strengthened the arguments of \cite{Colafrancesco2006} to explore the potential of radio emission as an indirect probe in clusters. Historically this has been difficult due to a lack of information about the magnetic field configuration in these environments~\citep{storm2017}, a similar difficulty to that in galaxy and dwarf galaxy searches~\cite{egorov2013,atca_II,chan2017,regis2017,Beck2019dsph,beckm312019,Chan_2019}. However, recent advances, such as the rotation measure study in \citet{Bonafede_2010}, have allowed the Coma cluster in particular to produce more powerful, and robust, indirect DM constraints from diffuse radio emissions~\citep{gs2016}. 
So far, something that has not been fully leveraged in DM searches is the fact that DM indirect radio emissions are highly non-thermal~\citep{gsp2015} and, moreover, have distinctly different energy spectra to the more mundane astrophysical power-law phenomena. The sense in which this property would be fully leveraged would be if one could find an indirect means of comparing the observed electron energy spectra producing non-thermal diffuse emission, in DM dominated environments, to the predicted electron spectra injected by DM annihilation/decays. The ideal scenario being a hard power-law spectrum inferred in a target environment contrasting with the far softer spectra that result from DM processes when the particle mass $\gtrsim \mathcal{O}(10)$ GeV. Such observations, however, can be most easily realised when a multi-frequency strategy is employed, as the convolution of electron and magnetic contributions must be disentangled. Once this is done, it is proposed here that the electron spectra comparison can be conducted indirectly through extrapolated synchrotron spectra produced under the same set of magnetic field assumptions. The synchrotron spectrum is used as a proxy as it's extrapolation towards spectral cut-off frequencies, where DM effects are most distinct from hard power-laws, can be probed by up-coming experiments like the next-generation Very Large Array (ngVLA). 

In \citet{Murgia_2010} the authors make use of both radio and X-ray data to constrain both the volume average of the magnetic field and the non-thermal electron population responsible for the observed radio mini-halo within $7^\prime$ of the centre of the Ophiuchus cluster. This is done by assumption that a single electron population produces both the non-thermal radio and X-ray emissions. This electron population is found to have a steep power-law slope $\sim 3.8$, which, combined with the volume averaged magnetic field strength of $0.3$ $\mu$G, suggests a synchrotron spectrum cutting off rapidly above $\mathcal{O}(1)$ GHz~\citep{Murgia_2010}. This can be supplemented by radial profile of the electron density in Ophiuchus found in \citet{Werner_2016}, which is well-fitted by a profile of the same form as found in \citet{1992A&A...259L..31B} for the Coma cluster. This similarity can be pushed one step forward by assuming the magnetic field profile will also follow the electron density, as found by \citet{Bonafede_2010} in the case of Coma. With these ingredients in hand one can compare the predicted DM radio emissions in Ophiuchus to only the synchrotron radiation from the non-thermal electron population. This is only possible when we can determine the electron population responsible for a given radio emission independent of magnetic field assumptions, which necessitates a multi-frequency approach. 
In this work we present the results of the indirect comparison of the energy-spectra of the observed electron population in the Ophiuchus mini-halo and that predicted for DM injection. This is facilitated through the drastically different synchrotron cut-off and peak frequencies for WIMP masses $\gtrsim \mathcal{O}(10)$ GeV. Importantly, for $\mu$G magnetic fields in galaxy clusters, discrimination of the synchrotron cut-off requires that measurements of radio emissions at frequencies above 10 GHz can be made on arcminute scales down to sensitivities of at least the $\mu$Jy level. We therefore demonstrate the indirect electron spectrum comparison at a level of sensitivity motivated by the up-coming ngVLA at frequencies below 100 GHz\footnote{\url{https://ngvla.nrao.edu/page/refdesign}}. This is done by extrapolating the synchrotron spectra to a maximum frequency chosen according to ngVLA observability. We demonstrate that, for a variety of potential magnetic field geometries, the constraints on the DM annihilation cross-section are far better than from existing VLA data and even exceed those from Fermi-LAT from dwarf galaxies~\cite{Fermidwarves2015,Fermidwarves2016} over a range of WIMP masses above $50$ GeV.

This work is structured as follows: section~\ref{sec:dm} discusses the calculation of DM radio emissions, \ref{sec:oph} details the modelling of Ophiuchus cluster environment, results are displayed in section~\ref{sec:res}, and finally conclusions are detailed in \ref{sec:conc}.



\section{DM radio emissions}
\label{sec:dm}
The source function  for electrons/positrons from DM annihilation is given by
\begin{equation}
Q_e (r,E) = \frac{1}{2}\langle \sigma V\rangle \sum\limits_{f}^{} \td{N^f_e}{E}{} B_f \left(\frac{\rho_{\chi}(r)}{m_{\chi}}\right)^2 \; ,
\end{equation}
where $\langle \sigma V\rangle$ is the velocity averaged annihilation cross-section, $f$ is the label of intermediate state (annihilation channel), $\td{N^f_e}{E}{}$ are the electrons per unit energy per annihilation produced found using \citet{ppdmcb1,ppdmcb2}, $B_f$ is the branching ratio, finally $\rho_{\chi}$ and $m_{\chi}$ are the WIMP density and mass respectively.

To calculate the synchrotron power per electron of energy $E$ we use~\citep{longair1994}
\begin{equation}
P_{synch} (\nu,E,r,z) = \int_0^\pi d\theta \, \frac{\sin{\theta}^2}{2}2\pi \sqrt{3} r_e m_e c \nu_g F_{synch}\left(\frac{\kappa}{\sin{\theta}}\right) \; ,
\label{eq:power}
\end{equation}
with
\begin{equation}
\kappa = \frac{2\nu (1+z)}{3\nu_g \gamma^2}\left[1 +\left(\frac{\gamma \nu_p}{\nu (1+z)}\right)^2\right]^{\frac{3}{2}} \; ,
\end{equation}
and
\begin{equation}
F_{synch}(x) = x \int_x^{\infty} dy \, K_{5/3}(y) \simeq 1.25 x^{\frac{1}{3}} \mbox{e}^{-x} \left(648 + x^2\right)^{\frac{1}{12}} \; .
\end{equation}

This is converted into a flux first using an emissivity:
\begin{equation}
j_{i} (\nu,r,z) = \int_{m_e}^{M_\chi} dE \, \left(\td{n_{e^-}}{E}{} + \td{n_{e^+}}{E}{}\right) P_{i} (\nu,E,r,z) \; ,
\label{eq:emm}
\end{equation}
and then integrating over the studied volume
\begin{equation}
S_{i} (\nu,z) = \int_0^r d^3r^{\prime} \, \frac{j_{i}(\nu,r^{\prime},z)}{4 \pi (D_L^2 + (r^\prime)^2)} \; ,
\label{eq:flux}
\end{equation}
with $D_L$ being the luminosity distance to the centre of the target.

To account for diffusion and energy-loss by electrons produced by annihilation we find the equilibrium solution to the equation
\begin{equation}
\begin{aligned}
\pd{}{t}{}\td{n_e}{E}{} = & \; \gv{\nabla} \left( D(E,\v{r})\gv{\nabla}\td{n_e}{E}{}\right) + \pd{}{E}{}\left( b(E,\v{r}) \td{n_e}{E}{}\right) + Q_e(E,\v{r}) \; ,
\end{aligned}
\end{equation}
where $D$ and $b$ are diffusion and energy-loss functions. We use the solution detailed in \cite{Colafrancesco2006} with the diffusion and loss functions given by
\begin{equation}
D(E) = D_0 \left(\frac{d_0}{1 \; \mbox{kpc}}\right)^{\frac{2}{3}} \left(\frac{\overline{B}}{1 \; \mu\mbox{G}}\right)^{-\frac{1}{3}} \left(\frac{E}{1 \; \mbox{GeV}}\right)^{\frac{1}{3}}  \; , \label{eq:diff}
\end{equation}
where $D_0 = 3.1\times 10^{28}$ cm$^2$ s$^{-1}$, $\overline{B}$ is the average magnetic field strength, and
\begin{equation}
\begin{aligned}
b(E) & = b_{IC} E^2 (1+z)^4 + b_{sync} E^2 \overline{B}^2 \; \\ & + b_{Coul} \overline{n} \left(1 + \frac{1}{75}\log\left(\frac{\gamma}{\overline{n}}\right)\right) + b_{brem} \overline{n} \left( \log\left(\frac{\gamma}{\overline{n}}\right) + 0.36 \right) \;,
\end{aligned}
\label{eq:loss}
\end{equation}
where $\gamma$ is the electron Lorentz factor, $\overline{n}$ is the average electron density, while $b_{IC}$, $b_{synch}$, $b_{col}$, and $b_{brem}$ are the inverse-Compton, synchrotron, Coulomb, and bremsstrahlung energy-loss factors. These are given, in units of $10^{-16}$ GeV s$^{-1}$, by $0.25$, $0.0254$, $6.13$, and $1.51$ respectively. In this case we consider inverse-Compton scattering off cosmic microwave background photons.

\section{Ophiuchus cluster environment}
\label{sec:oph}

\subsection{Electron distribution}
We make use of Chandra observations~\citep{Werner_2016} in our calculation of the spatial distribution of the electron population within Ophiuchus. To the data of \citet{Werner_2016} we fit the following radial profile motivated by studies of the Coma cluster~\citep{1992A&A...259L..31B}
\begin{equation}
n_e (r) = n_0 \left(1+\frac{r^2}{r_c^2}\right)^\beta \; ,
\end{equation}
where $n_0 = 0.29 \pm 0.015$ cm$^{-3}$, $r_c = 0.99 \pm 0.12$ kpc, and $\beta = -0.41 \pm 0.012$ are found by fitting the data. We therefore use $\overline{n} = 0.017$ cm$^{-3}$ as the average electron density (note that this average was found within a $7^\prime$ radius and was weighted by the square of the DM density to better reflect the environment for the bulk of annihilations).

\subsubsection{Non-thermal electrons}
We will follow \citet{Murgia_2010} in their finding that the non-thermal electron population, in the energy range $\gamma = 300$ and $\gamma = 3\times 10^4$, has a power-law slope of $p = 3.83^{+0.38}_{-0.44}$ with normalisation $k_0 = 0.53^{+6.16}_{-0.52}$ cm$^{-3}$. In the interests of seeing the interaction of the spatial profiles of magnetic field and electron density we will further assume this population follows a spatial distribution of the same shape as above. This spatial distribution will be normalised so that its volume average over a radius of $7^\prime$ matches the power-law amplitude $k_0$.

\subsection{Magnetic field}
The radio study of Ophiuchus in \citet{Murgia_2010} determined a volume-averaged magnetic field strength of $0.3^{+0.11}_{-0.07}$ $\mu$G within a radius of $7^\prime$ from the cluster centre. We will assume the magnetic field follows a similar profile to that in Coma cluster~\citep{Bonafede_2010}
\begin{equation}
B (r) = B_0 \left(\frac{n_e(r)}{n_0}\right)^\eta \; .
\end{equation}
We note that with only a single data point from the volume average we can only constrain either $B_0$ or $\eta$ and leave the other free. For the purposes of calculating $\overline{B}$ we find the volume average within a $7^\prime$ radius weighted by the square of the DM density to better reflect the environment for the bulk of annihilations (this is important as functions $D$ and $b$ have no spatial dependence in the solution we employ).

\subsection{Chosen environmental values}
\begin{figure}
	\centering
	\resizebox{0.99\hsize}{!}{\includegraphics{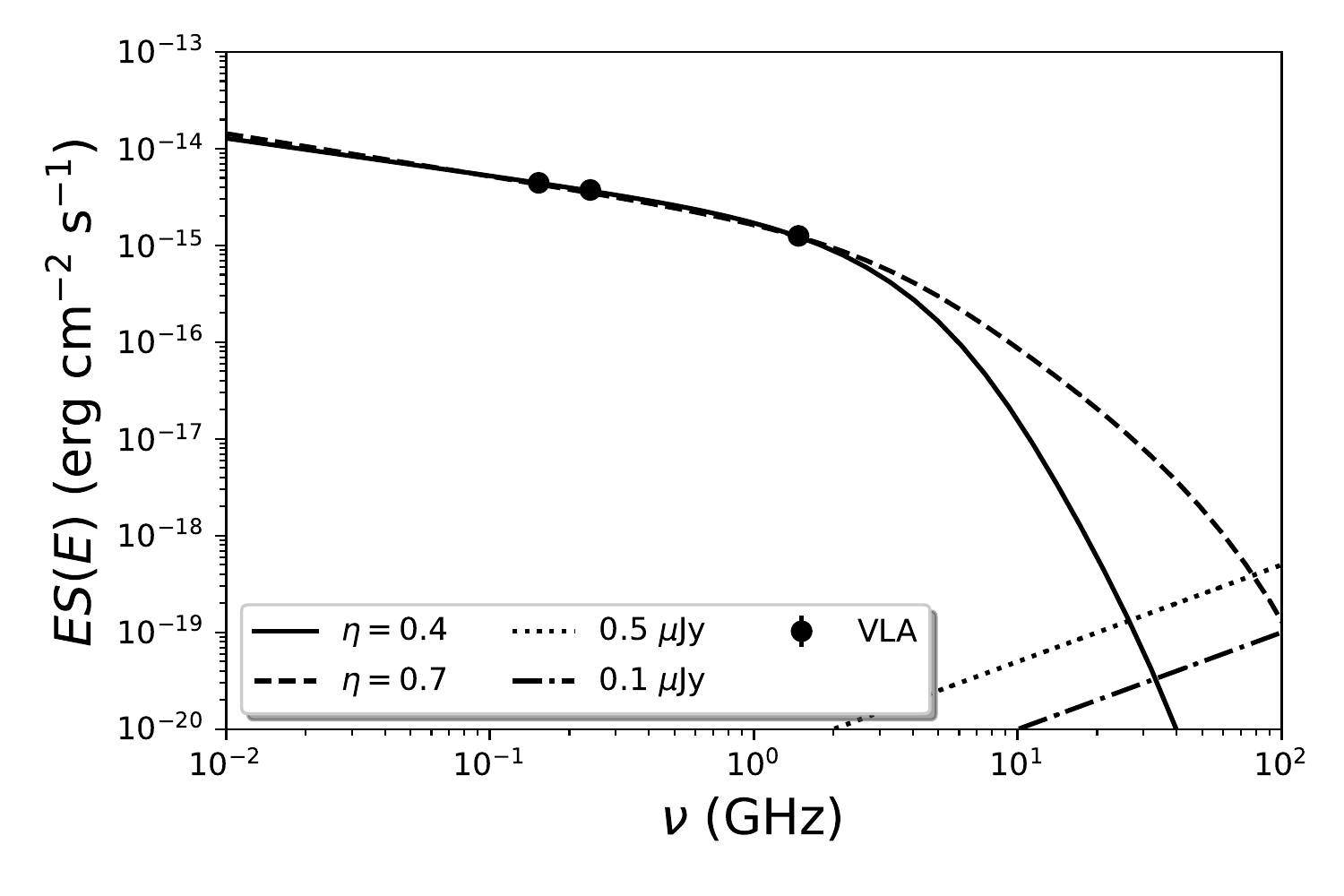}}
	\caption{Ophiuchus cluster mini-halo spectral extrapolations from the data points of \citet{Murgia_2010} (labelled as VLA in the key) without any DM effects included. The solid and dashed lines show the cases with $\eta = 0.4$ and $\eta =0.7$ respectively. The dotted and dash-dotted lines show the $0.5$ and $0.1$ $\mu$Jy thresholds.}
	\label{fig:spectrum}
\end{figure}

We elect to allow $\eta$ to take values in the range allowed in Coma~\citep{Bonafede_2010} and then choose $k_0$, $B_0$, and $p$ (from the allowed range for Ophiuchus) to best fit the mini-halo spectrum from \citet{Murgia_2010}. The values we find are tabulated in Table~\ref{tab:eta}. The spectral extrapolations with sensitivity levels are displayed in Figure~\ref{fig:spectrum}.
\begin{table}
	\centering
	\caption{Environmental variable choices. $\eta$ is allowed to range freely across the values found for the Coma cluster in \citet{Bonafede_2010} and the others are found as best fits within the ranges from \citet{Murgia_2010}.}
	\label{tab:eta}
	\begin{tabular}{|l|l|l|l|}
		\hline
	 	$\eta$ & $B_0$ ($\mu$G) & $k_0$ (cm$^{-3}$) & $p$ \\
	 	\hline
	 	0.4 & 1.4 & 0.15 & 3.8 \\
	 	0.7 & 5.2 & 0.24 & 3.9 \\
	 	\hline
	\end{tabular}
\end{table}

\subsection{DM halo}
In this work we will derive the parameters of the dark matter halo in Ophiuchus from optical studies performed in \citet{Durret_2015}. Thus, we use $M_{vir} = 1.1 \times 10^{15}$ M$_\odot$, $r_{vir} = 2.1$ Mpc, with a Navarro-Frenk-White (NFW) density profile~\citep{nfw1996}
\begin{equation}
\rho_{nfw}(r)=\frac{\rho_{s}}{\frac{r}{r_s}\left(1+\frac{r}{r_s}\right)^{2}} \; , 
\label{eq:density}
\end{equation}
with $\rho_s$ set by normalisation and $r_s = 0.7$ Mpc.

Importantly, we use no form of substructure boosting factor in this work.

\section{Results}
\label{sec:res}
Since DM radio emissions are highly non-thermal~\citep{gsp2015} we propose in this work to derive constraints on DM annihilation by comparing only to the synchrotron emission sourced from the non-thermal electron population in the target object. In the case of Ophiuchus \citet{Murgia_2010} determined this by assuming a single power-law population was responsible for the synchrotron and X-ray emissions. We will use their non-thermal population and calculate the expected synchrotron emissions and then compare this to the DM predictions. 

 \begin{figure}
	\centering
	\resizebox{0.99\hsize}{!}{\includegraphics{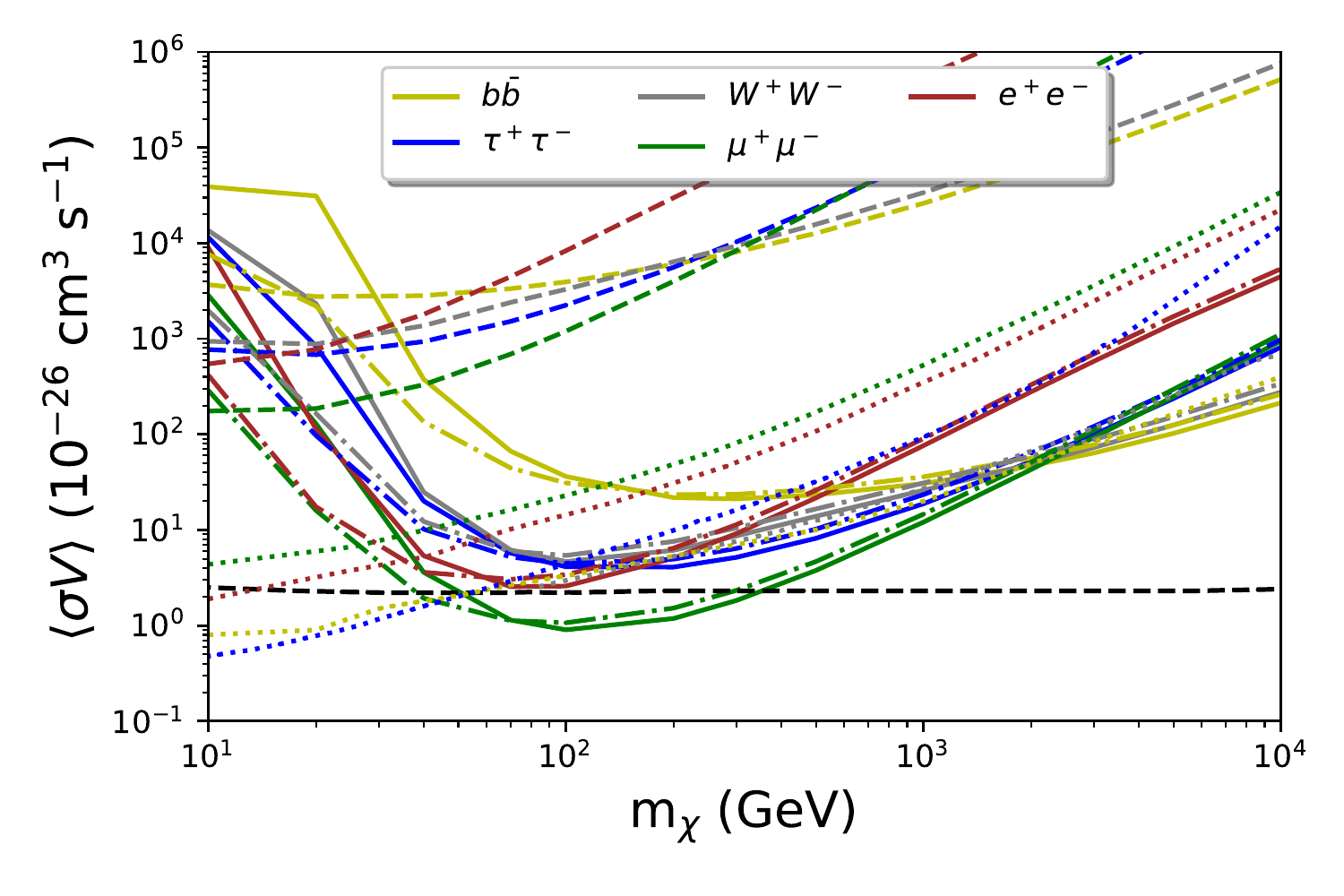}}
	\caption{Ophiuchus cluster limits on $\langle \sigma V\rangle$ as a function of WIMP mass $m_\chi$ assuming minimum observable flux of $0.5$ $\mu$Jy below 100 GHz. Solid lines show $\eta = 0.4$ and $B_0 = 1.4$ $\mu$G while dot-dashed lines show $\eta = 0.7$ and $B_0 = 5.2$ $\mu$G. The black dashed line shows the thermal relic cross-section from \citet{steigman2012}. Other dashed lines show constraints derived from the VLA data points~\citep{Murgia_2010}. The displayed Fermi limits (dotted lines) come from \citet{Fermidwarves2015,Fermidwarves2016}.}
	\label{fig:oph1}
\end{figure}

In Figure~\ref{fig:oph1} we display the constraints found by comparison of the non-thermal synchrotron emissions and DM predictions. This figure covers the range of magnetic field models that we have studied. The case we display extrapolates the Ophiuchus mini-halo spectrum up to frequencies where DM predictions change the synchrotron cut-off, the maximum frequency chosen is such that the DM flux is in $2\sigma$ excess above the mini-halo extrapolation, which has a minimum flux of $ \gtrsim 0.5$ $\mu$Jy at the maximum extrapolation frequency. This minimum flux is chosen to be similar to the 1 hour continuum rms per beam of the ngVLA baseline design\footnote{\url{https://ngvla.nrao.edu/page/refdesign}}, which ranges between 0.2 and 0.7 $\mu$Jy over the ngVLA frequency bands.   

The principle feature of Fig.~\ref{fig:oph1} is that above $30$ GeV WIMP masses the results found here exceed those drawn from the existing VLA data points~\citep{Murgia_2010} by around 2 orders of magnitude in all annihilation channels. Additionally, these results better Fermi-LAT dwarf searches for WIMP masses $> 50$ GeV in all but the $b$-quark and $W$-boson channels. In the lepton channels our results improve on Fermi-LAT by around an order of magnitude. In the case of $b\bar{b}$ and $W^+W^-$ these achieve around a factor of $2$ better than Fermi-LAT at masses above $1$ TeV. It is worth noting that, with our use of a radial profile for magnetic field and electron distributions, the synchrotron cut-off is less severe than in the fitting done by \citet{Murgia_2010}, so this aspect of our modelling is not making the results unduly optimistic. Importantly, given the chosen minimum detectable flux, very similar results are achieved across the range of magnetic field models studied.

 \begin{figure}
	\centering
	\resizebox{0.99\hsize}{!}{\includegraphics{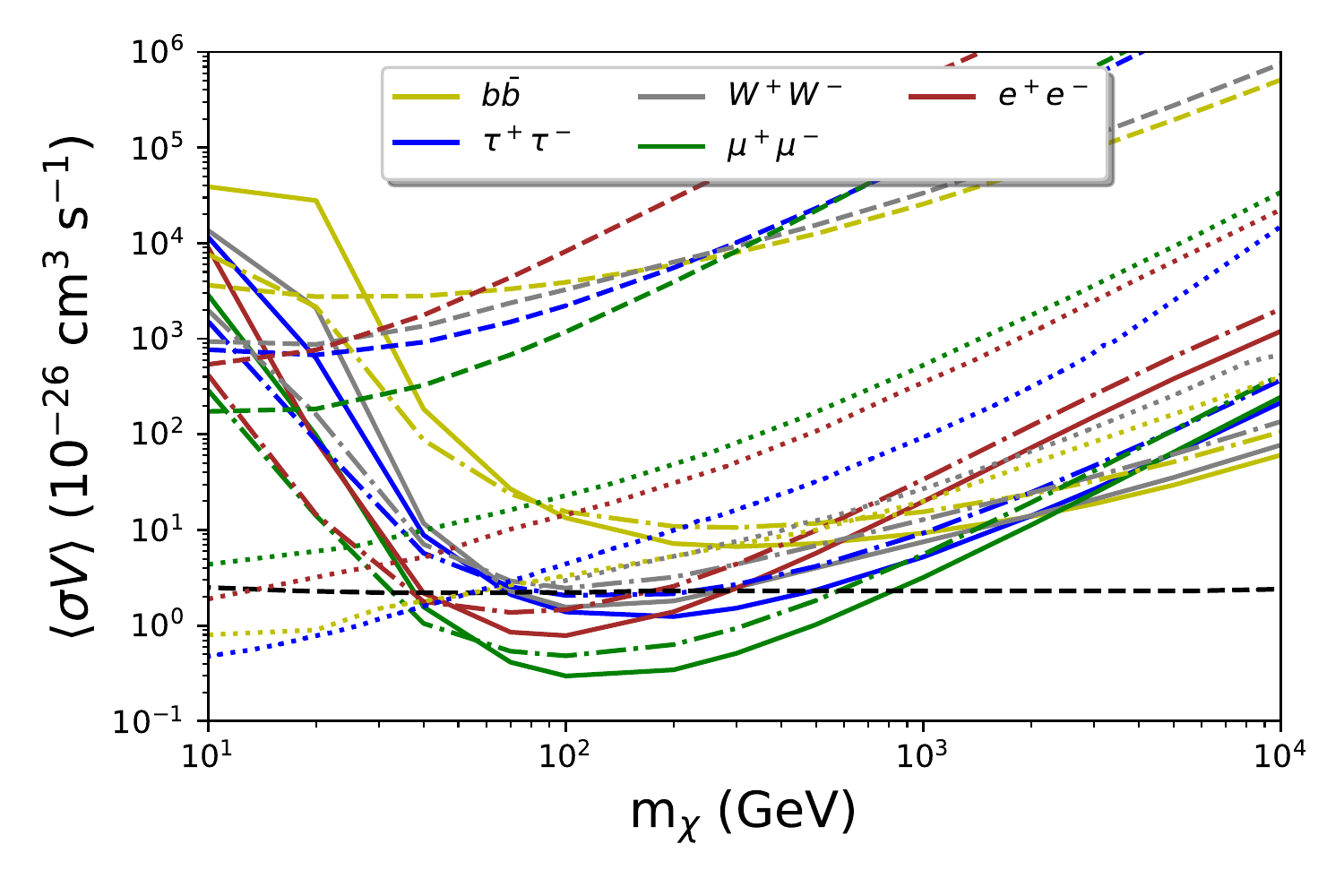}}
	\caption{Ophiuchus cluster limits on $\langle \sigma V\rangle$ as a function of WIMP mass $m_\chi$ assuming minimum observable flux of $0.1$ $\mu$Jy below 100 GHz. Solid lines show $\eta = 0.4$ and $B_0 = 1.4$ $\mu$G while dot-dashed lines show $\eta = 0.7$ and $B_0 = 5.2$ $\mu$G. The black dashed line shows the thermal relic cross-section from \citet{steigman2012}. Other dashed lines show constraints derived from the VLA data points~\citep{Murgia_2010}. The displayed Fermi limits (dotted lines) come from \citet{Fermidwarves2015,Fermidwarves2016}.}
	\label{fig:oph2}
\end{figure}

In Figure~\ref{fig:oph2} we display a similar set of results to Fig.~\ref{fig:oph1} but we assume a minimum flux of $0.1$ $\mu$Jy as observable. The effects on the weaker central magnetic field case are stronger as the extrapolation falls off more dramatically at lower frequencies (see Fig.~\ref{fig:spectrum}), producing a more significant clash with the softer DM spectrum. However, the effect on both scenarios is substantial, improving the constraints by nearly a factor of $5$ and extending the range in which the $b\bar{b}$ and $W^+W^-$ exceed Fermi-LAT limits to above $200$ GeV WIMP masses. The fact that the weaker magnetic field can produce stronger constraints is remarkable in the study of indirect DM radio emissions and will be remarked upon further below.

This brings us to the discussion of both the strong potential of this constraint approach and why it responds as it does to the choice of magnetic field geometry. The latter point can be addressed by noting the steepness of the power-law slope that characterises the non-thermal emitting electron population, being around $\sim 3.8$. This is very different to energetics of electrons from DM-based injection~\cite{ppdmcb1,ppdmcb2} which tend to be much flatter and cut-off around the WIMP mass. This means that a non-thermal population produced by DM injection will likely have a far softer spectrum in energy. Thus, with a given magnetic field, both the synchrotron peak and cut-off will occur at higher frequencies for the DM case than the power-law (for WIMP masses above a few tens of GeVs). The weaker magnetic field means that the discrepancy becomes noticeable at a lower frequency. With the stronger field, however, the peak of the DM distribution rapidly shifts out of the window for considering the Ophiuchus mini-halo spectrum (we use a maximum possible extrapolation limit of 100 GHz to match the ngVLA and avoid considering emissions beyond that of synchrotron). The reason the discrepancy appears only at higher radio frequencies is the steepness of the inferred electron population which favours far stronger synchrotron emissions at lower frequencies, and these greatly exceed DM predictions. The mass dependence of Figs.~\ref{fig:oph1} and \ref{fig:oph2} is also explained in this manner: the larger the WIMP mass the higher the electron energy cut-off and consequently the softer the synchrotron spectrum between 10 MHz and 100 GHz (note that for larger WIMP masses $\gtrsim 1$ TeV the synchrotron slope is positive in the studied frequency range) while the power-law derived case remains with a hard cut-off above $\mathcal{O}(1)$ GHz.

Although the technique used here is comparing the mismatch of the electron populations indirectly, it is only searching for excess emissions above the synchrotron spectrum predicted from the power-law distribution found in \citet{Murgia_2010}. In addition, by fitting to the mini-halo spectrum from \citet{Murgia_2010} and limiting our extrapolation to a range explorable with the ngVLA we ensure that, although we are not directly comparing DM predictions to extant data points, our results are strongly grounded in empirical data. 

The question of whether the presence of a second harder power-law population of electrons would affect the DM limits is an important one to consider. It is vital to first note that \citet{Murgia_2010} study a range of electron $\gamma$ factors for their population that maintains consistency between radio and X-ray data and that the limits of this range are of vital importance in the frequency position of the synchrotron cut-off. Thus, the presence of a secondary harder population within this energy range would be limited to adjusting the continuation of the synchrotron cut-off above 1.5 GHz. Since the magnetic field and electron population energy range are somewhat degenerate in their effect on the synchrotron cut-off frequency we would then expect the magnitude of DM limit uncertainty due to this possibility to be similar to that from the magnetic field profile (which is seen to be rather small in Figs.~\ref{fig:oph1} and \ref{fig:oph2}). This suggests that the results presented here would not be strongly sensitive to the presence of a broken power-law electron population rather than a single one.

An additional source of uncertainty to consider here is the diffusive environment model being used. In \citet{Colafrancesco2006} and \citet{gsp2015} it was established that larger structures like galaxy clusters tend to have their DM annihilation synchrotron flux dominated by energy-loss effects rather than diffusion. This is because the length scales under consideration tend to be much larger than the diffusive scale. In this case we consider a $7^\prime$ angular radius region of Ophiuchus which is still $\approx 460$ kpc in extent. Since the typical diffusion scales for DM-produced electrons is around the kpc range~\citep{Colafrancesco2006} we should expect that the energy-loss uncertainties are significantly more important than those from diffusion. The most substantial energy-loss uncertainties are the magnetic field strength and the energy density of target photons for inverse-Compton scattering~\cite{egorov2013}. In this work we have attempted to make the energy-loss assumptions conservative by using a magnetic field average strength weighted by $\rho_{DM}^2$ so that it is more sensitive to steepness of the magnetic field profile. In the inverse-Compton case it is unclear, in a galaxy cluster environment, if an additional photon population should be considered beyond the CMB. However, in the case of galaxy environments the impact of an inter-stellar radiation field can be significant~\cite{egorov2013}.

\section{Conclusions}
\label{sec:conc}

We have demonstrated the potential of indirect DM probes that rely upon the comparison between the shapes of observed and predicted electron energy spectra. This is done using the Ophiuchus cluster where combined X-ray and radio data were used in \citet{Murgia_2010} to determine a non-thermal electron population responsible for observed multi-frequency emissions. Our results make use of spatial electron distributions, drawn from \citet{Werner_2016}, which also inform our magnetic field profiles following \citet{Bonafede_2010}, but we stress that this does not strongly affect our conclusions (indeed the results are slightly weaker than when just considering the volume averaged electron spectrum). In addition, because we consider the same magnetic field geometry (which was required to fit the mini-halo spectrum) applied to both observed and predicted electron populations we ensure the ability to actually compare electron spectra and also reduce the relative magnetic field uncertainty present in radio-frequency indirect DM searches. We note, however, that diffusive uncertainties will still have their full impact and we chose our diffusion scenario to produce conservative results. 

The results we displayed showed that over a range of magnetic field profile choices we can produce results superior to the Fermi-LAT searches of dwarf galaxies~\cite{Fermidwarves2015,Fermidwarves2016} over a very wide range of WIMP masses. Interestingly, our results have the potential to improve when the central field strength is weaker and we argue that this is because we are able to fully leverage the mismatch in synchrotron peak and cut-off frequencies in these cases. Importantly, for central field strengths below $5$ $\mu$G, we can probe below the thermal relic cross-section for WIMP masses around 100 GeV in all the considered annihilation channels barring $b$-quarks. The improvement over Fermi-LAT is most marked at larger WIMP masses and for the leptonic annihilation channels, where order of magnitude advantages are obtained even in the worst-case magnetic field geometry. We stress that, although this technique depends on extrapolated synchrotron spectra, we have limited our extrapolation to frequencies and minimum fluxes that can be potentially probed by the ngVLA.

\section*{Acknowledgements}
G.B acknowledges support from a National Research Foundation of South Africa Thuthuka grant no. 117969. This research has made use of the NASA/IPAC Extragalactic Database (NED), which is operated by the Jet Propulsion Laboratory, California Institute of Technology, under contract with the National Aeronautics and Space Administration. This work also made use of the WebPlotDigitizer\footnote{\url{http://automeris.io/WebPlotDigitizer/}}.




\bibliographystyle{mnras}

\bibliography{ophiuchus,dm_indirect_general,dm_indirect,dsph_all,gbeck}





\bsp	
\label{lastpage}
\end{document}